\documentclass[aip,graphicx]{revtex4-1}

\draft 

\usepackage{graphicx}
\usepackage{dcolumn}
\usepackage{bm}

\usepackage[utf8]{inputenc}
\usepackage[T1]{fontenc}
\usepackage{mathptmx}
\usepackage{etoolbox}

\makeatletter
\def\@email#1#2{%
 \endgroup
 \patchcmd{\titleblock@produce}
  {\frontmatter@RRAPformat}
  {\frontmatter@RRAPformat{\produce@RRAP{*#1\href{mailto:#2}{#2}}}\frontmatter@RRAPformat}
  {}{}
}%
\makeatother

\begin{document}


\title{Terahertz emission from transient currents and coherent phonons \\in layered MoSe$_2$ and WSe$_2$} 



\author{Jessica Afalla}
\email[]{afalla.castillo.gn@u.tsukuba.ac.jp}
\affiliation{Faculty of Pure and Applied Sciences, University of Tsukuba, Tennodai 1-1-1 Tsukuba 305-8573 Japan}

\author{Joselito Muldera}
\affiliation{Research Center for Development of Far Infrared Region, University of Fukui, Bunkyo 3-9-1 Fukui 910-8507 Japan}

\author{Semmi Takamizawa}
\affiliation{Graduate School of Pure and Applied Sciences, University of Tsukuba, Tennodai 1-1-1 Tsukuba 305-8573 Japan}

\author{Takumi Fukuda}
\affiliation{Graduate School of Pure and Applied Sciences, University of Tsukuba, Tennodai 1-1-1 Tsukuba 305-8573 Japan}

\author{Keiji Ueno}
\affiliation{Department of Chemistry, Saitama University, Shimookubo 255 Saitama 338-8570, Japan}

\author{Masahiko Tani}
\affiliation{Research Center for Development of Far Infrared Region, University of Fukui, Bunkyo 3-9-1 Fukui 910-8507 Japan}

\author{Muneaki Hase}
\affiliation{Faculty of Pure and Applied Sciences, University of Tsukuba, Tennodai 1-1-1 Tsukuba 305-8573 Japan}


\date{\today}

\begin{abstract}
Terahertz (THz) time-domain emission spectroscopy was performed on layered 2H-MoSe$_2$ and 2H-WSe$_2$. The THz emission shows an initial cycle attributed to surge currents and is followed by oscillations attributed to coherent interlayer phonon modes. To obtain the frequencies of the interlayer vibrations, analysis of the THz emission waveforms were performed, separating the two contributions to the total waveform. Results of the fitting show several vibrational modes in the range of 5.87 – 32.75 cm$^{-1}$ for the samples, attributed to infrared-active interlayer shear and breathing modes. This study demonstrates that THz emission spectroscopy provides a means of observing these low frequency vibrational modes in layered materials.
\end{abstract}


\maketitle 

\section{Introduction}
Transition metal dichalcogenides (TMDCs) are a family of layered materials forming MX$_2$ (M=transition metal, X=chalcogen), with a wide range of electronic, mechanical, and optical properties\cite{wang2012electronics}. One notable feature that attracted interest in the material is the indirect-to-direct transition of the bandgap as the TMDC layer thickness reaches monolayer level\cite{splendiani2010emerging}, wherein the monolayer bandgap coincides with the $A$-gap found in the TMDC's band structure K-point. Among the different TMDCs, hexagonal 2H-MoSe$_2$ and 2H-WSe$_2$ are semiconducting, with bandgaps falling within the range of typical femtosecond laser wavelengths. Monolayer MoSe$_2$ has a bandgap around 1.57 eV\cite{tonndorf2013photoluminescence} and an indirect bulk bandgap 1.122 eV \cite{kam1984fundamental}, while monolayer WSe$_2$ has a bandgap around 1.6 eV\cite{tonndorf2013photoluminescence,zhao2013origin}, and an indirect bulk bandgap of 1.21 eV \cite{tonndorf2013photoluminescence,kam1984fundamental}. Having a large on/off ratio, ultrafast speed, and high photoresponsivity, semiconducting TMDCs have great potential for ultrafast optical devices\cite{shi2022c}, for which prototype THz modulators, photodetectors, electrical energy storage\cite{fan2020terahertz}, and many others\cite{fan2020terahertz,fan2020characteristics} have been reported. 

In TMDCs, each MX$_2$ layer is only weakly bonded to the next via van der Waals interaction, which makes the mechanical exfoliation into single layers possible \cite{ueno1990heteroepitaxial,azizi2015freestanding}.  The properties of few-layer TMDCs, including bandgap, frictional characteristics, and electron-phonon interactions, are highly dependent on thickness and layer number\cite{zhao2013interlayer}. Interlayer interactions can have detrimental effects on device performance, and is gaining attention in recent years \cite{sie2019ultrafast,mannebach2017dynamic}. Interlayer vibration refers to the movement of each layer of MX$_2$ as a single unit, with the weak van der Waals interaction acting as the restoring force. Due to this nature, interlayer vibrations occur at low frequencies, such as in the tens of wavenumbers. The low phonon energy also make these lattice vibrations difficult to detect\cite{ge2014coherent}. Reports using modified Raman detection schemes successfully predicted and measured Raman-active shear and layer breathing modes in few- to multi-layer TMDCs\cite{zhao2013interlayer,liang2017low,zhang2015phonon}. For typical Raman spectroscopies (e.g. measurement limit $\sim$10 cm$^{-1}$) however, these “ultralow” frequencies \cite{zhang2015phonon} are in close proximity to the laser line\cite{taday2003using}.

An alternative approach is using ultrafast excitation. In certain semiconductors, ultrafast excitation produces THz dipole radiation due to the effective separation of electrons and holes either via drift or diffusion (transient current)\cite{sakai2005introduction,gu2005terahertz}. The same separation of charges can also screen the surface field and can launch coherent lattice vibrations\cite{pfeifer1992generation,dekorsy1995emission,tani1998terahertz,ge2014coherent}. Semiconductors such as Ge, Te, and GaAs have been shown to emit THz radiation from both transient currents as well as from infrared(IR)-active coherent LO and TO phonons\cite{pfeifer1992generation,dekorsy1995emission,tani1998terahertz}. In a report on Te and CdTe\cite{tani1998terahertz}, the measured THz emission was described as having an initial radiation burst followed by damped oscillations. The initial burst, in the form of a single cycle pulse, originates from the separation of charges due to the induced photo-Dember field, while the damped oscillations are attributed to coherent phonon vibrations. Additionally, measuring in the time-domain also provides information on the lifetime of the phonons detected. In this work, we measured THz emission from multilayer MoSe$_2$ and WSe$_2$. The time-domain waveforms show an initial cycle pulse which are immediately followed by highly damped oscillations. We attribute the initial cycle pulse to the surge current effect, and the latter to interlayer phonon vibrations which can occur due to the weak interlayer bonding in TMDCs. 

\section{Methods}
For this report, bulk single crystal 2H-MoSe$_2$ and 2H-WSe$_2$ were grown using physical vapor transport method\cite{ueno2015introduction,lieth1977preparation}. The method yields free standing, millimeter-sized single crystal samples that are nominally defect-free. Small pieces were directly harvested from quartz tubes used during growth, and lightly adhered to aluminum sample holders using double-sided tape. The adhered samples were pristine, i.e. no exfoliation was performed, with thicknesses in the range of tens of micrometers.

We performed THz time-domain emission spectroscopy (THz-TDES) using a mode-locked Ti:sapphire laser with $\sim$100 fs pulse duration, operating at a repetition rate of 82 MHz. The output wavelength was varied between 765-830 nm. The samples were optically excited under a maximum fluence of $\sim$0.103 $\mu$J/cm$^2$ (spot size=0.9 mm); the pump beam was modulated at 2 kHz by an optical chopper, loosely focused on the sample surface, and incident at $\sim$45$^{\circ}$. The emitted THz radiation was detected by a LT-GaAs photoconductive dipole antenna, with a gap of 3.4 $\mu$m. The average power of the probe beam triggering the antenna was fixed at 10 mW. Experiments were conducted in open-air laboratory conditions. 

\section{Results and Discussion}
\begin{figure}
\includegraphics[width=0.6\columnwidth]{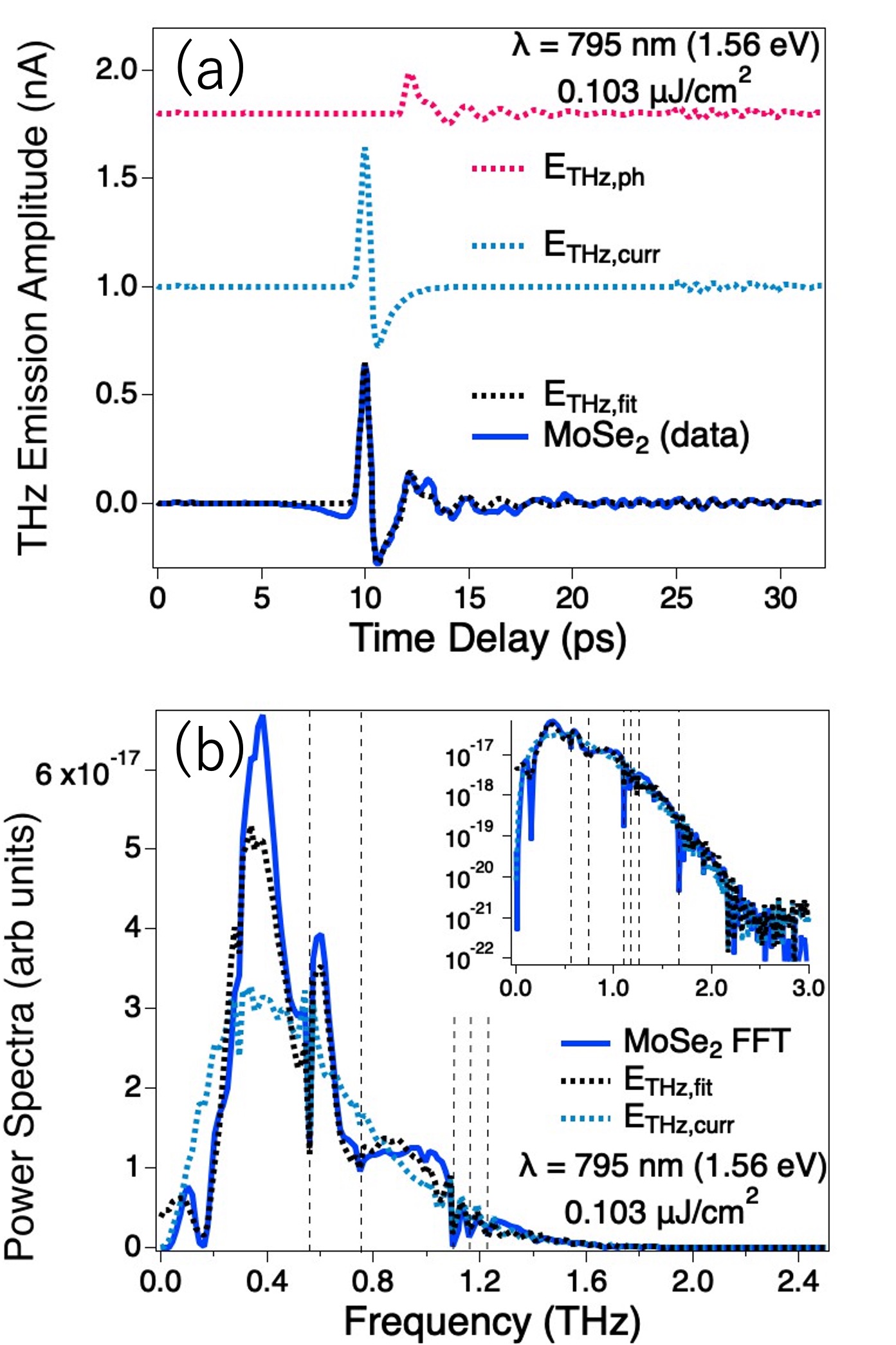} 
\caption{\label{MoSe2data} (a) Time-domain THz emission waveform of MoSe$_2$, blue solid curve shows the measured data, the black dashed curve is the full waveform fit. The light blue dashed curve represents the current-driven emission $E_{THz,curr}$ and the pink dashed curve is the phonon-driven THz emission $E_{THz,ph}$. The corresponding frequency spectra are shown in linear scale in (b), the inset shows the same curves in semilogarithmic scale. Vertical lines mark the water vapor lines.}
\end{figure}

THz emission from single crystal MoSe$_2$ excited at 795 nm (p-polarized, with fluence of 0.103 $\mu$J/cm$^2$) is shown in Fig.\ref{MoSe2data}, along with the corresponding Fourier transformed spectra. The time-domain waveform is shown at the bottom of Fig.1(a) as a blue solid curve. It has a single cycle pulse, with the peak centered around time delay = 10 ps, followed by oscillatory features. The oscillatory features are a combination of what is posited to be emission from interlayer phonons, and signals attributed to water vapor. The measured data is shown overlapped with a waveform fit ($E_{\text{THz,fit}}$), which can be separated into the single cycle pulse $E_{\text{THz,curr}}$ and the phonon-attributed THz emission, $E_{\text{THz,ph}}$, which lies between time delay = 10 to 15 ps. The attributions of the THz emission, along with the details of the fitting, will be discussed later in this section. The rest of the persistent oscillations are attributed to water vapor\cite{VanExter1989}, as the experiments were done in open air, laboratory conditions.  The frequency spectra of the measured emission is shown as a blue solid curve in Fig.1(b), overlaid with the Fourier-transform of the waveform fit. The Fourier-transform of the single cycle pulse $E_{\text{THz,curr}}$ is presented in the plot as well (light blue dashed curve). Comparing the power spectra of the full waveform ($E_{\text{THz,fit}}$) with that of $E_{\text{THz,curr}}$, spectral features in the form of peaks around 0.38 THz and 0.6 THz, as well as spectral dips around 0.2 and 0.7 THz become apparent. These features are still observable in the semi-logarithmic plot shown as an inset. To distinguish frequency features attributed to water vapor, vertical dashed lines are used to mark the prominent dips at 0.56, 0.74, 1.10, 1.16, 1.23 THz.

THz emission from optically excited surfaces can typically come from optical rectification in nonlinear media giving rise to a nonlinear polarization $P$, or from the dipole field radiation produced by the effective separation of charge carriers in semiconductors\cite{sakai2005introduction} having a time-varying current density $J$. In cases where the material is semiconducting and has a non-centrosymmetric crystal structure, THz emission arises as a combination of both phenomena. The time-domain waveform generated from these mechanisms can be described by the first two terms on the right-hand side of Eq. (1),
\begin{equation}
 E_{\text{THz}}\propto\frac{\partial^2 P}{\partial t^2}+\frac{\partial J}{\partial t}+\text{others} 
\end{equation}
The third term  (“others”) encompasses other means by which THz radiation is produced, such as THz radiation from coherent phonons\cite{tani1998terahertz,dekorsy1995emission,pfeifer1992generation}, spintronic emission\cite{papaioannou2020thz}, and other phenomena. For our results, the low frequencies of the oscillatory features suggest that these originate from infrared-active coherent interlayer phonon vibrations, which can also create time-varying changes in the crystal dipole moment. In this work, we attempt to obtain the frequencies of these phonon modes by analyzing the time-domain waveforms. And prior to this process, we first establish the THz generation mechanism from our samples. 

THz emission has been reported for hexagonal TMDCs, such as MoS$_2$ and WS$_2$\cite{zhang2017terahertz, huang2018terahertz, nevinskas2021terahertz}, MoSe$_2$\cite{nevinskas2021terahertz, yagodkin2021ultrafast} and WSe$_2$\cite{nevinskas2021terahertz, gorbatova2019terahertz}, differing with each type of TMDC. For the widely-studied MoS$_2$, azimuthal angle experiments reveal that nonlinear optical effect is the dominant mechanism, with secondary contributions from transient current\cite{zhang2017terahertz,huang2018terahertz}; meanwhile, in WS$_2$, the nonlinear optical effect is reported to be less pronounced\cite{zhang2017terahertz}. Only a few reports could be found for the THz emission of WSe$_2$. Using below-bandgap excitation, THz emission was observed\cite{gorbatova2019terahertz} in monolayers and therefore attributed to nonlinear optical effects; however, measurements on bulk WSe$_2$ above bandgap showed no azimuthal angle dependence and consequently attributed to the transient currents\cite{nevinskas2021terahertz}. For TMDCs with 2H stacking, bulk crystals have inversion symmetry and belong to D$_{6h}$ space group\cite{wen2019nonlinear}. Inversion symmetry breaking, and thus nonlinear polarization effects, are typically only induced at the surface, which becomes important for few to single layer samples. For MoSe$_2$, a report on the THz emission from both CVD-grown bulk and monolayer samples showed that it is combination of the nonlinear optical effect and the transient current effect, with the latter having the higher contribution\cite{yagodkin2021ultrafast}. Furthermore, they observed in-plane currents and features attributed to quantum beats. In a report of THz experiments done using similar bulk samples of MoSe$_2$ and WSe$_2$ as ours, Nevinskas et al attributes the THz emission solely from surface fields giving rise to the transient currents\cite{nevinskas2021terahertz}. 

To clarify the origin of the initial THz pulse for our bulk samples, we conducted azimuthal angle-dependence measurements on MoSe$_2$. If the THz emission is due to nonlinear optical effects, we should expect THz emission that is dependent on the crystal axis orientation. By exciting the sample by a p-polarized pump beam, the x- and y-components of the emitted THz electric field were measured separately, using a pair of wire grid polarizers (Fig.2(a)). A 3D plot of the time-domain waveforms is shown in Fig.2(b) for different azimuths. As the sample is rotated from its original position up to 90º, the THz emission remains p-polarized, emphasized by the $E_{THz,x}$ vs $E_{THz,y}$ plot in Fig.2(c). Although some changes in the $E_{THz,x}$ amplitude can be observed, we attribute these to slight variations in the illuminated spot as the sample is rotated. 

\begin{figure*}
\includegraphics[width=1.0\columnwidth]{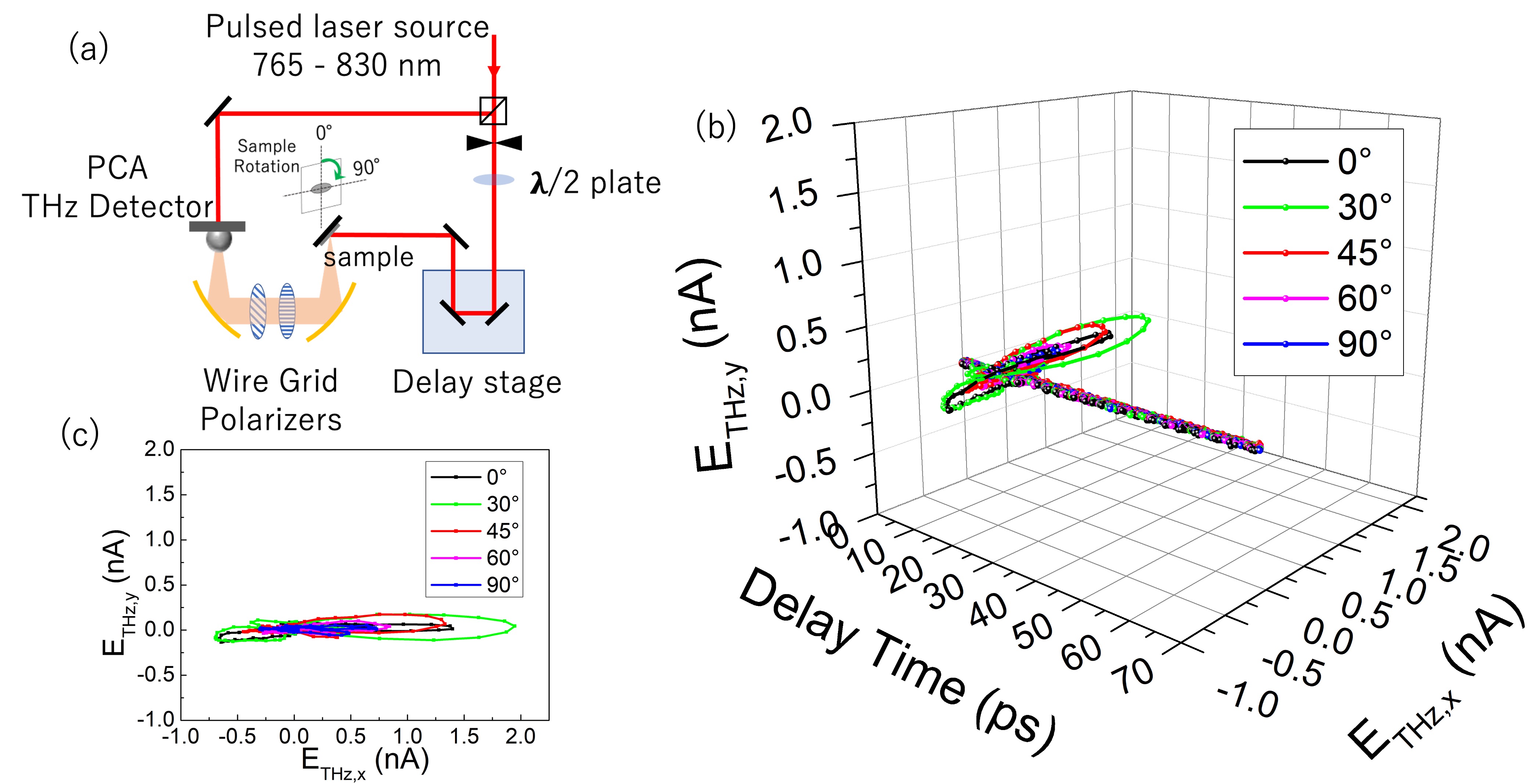}
\caption{\label{azimuthal} Azimuthal angle-dependent measurements of MoSe$_2$ emission under 795 nm excitation. (a) schematic of the experiment, (b) 3D plot showing the x- and y- components of the time-domain THz electric field as the crystal is rotated 90º, (c) $E_{\text{x,THz}}$ vs $E_{\text{y,THz}}$ plot emphasizing the invariance of the p-polarized emission.}
\end{figure*}

\begin{figure}
\includegraphics[width=1.0\columnwidth]{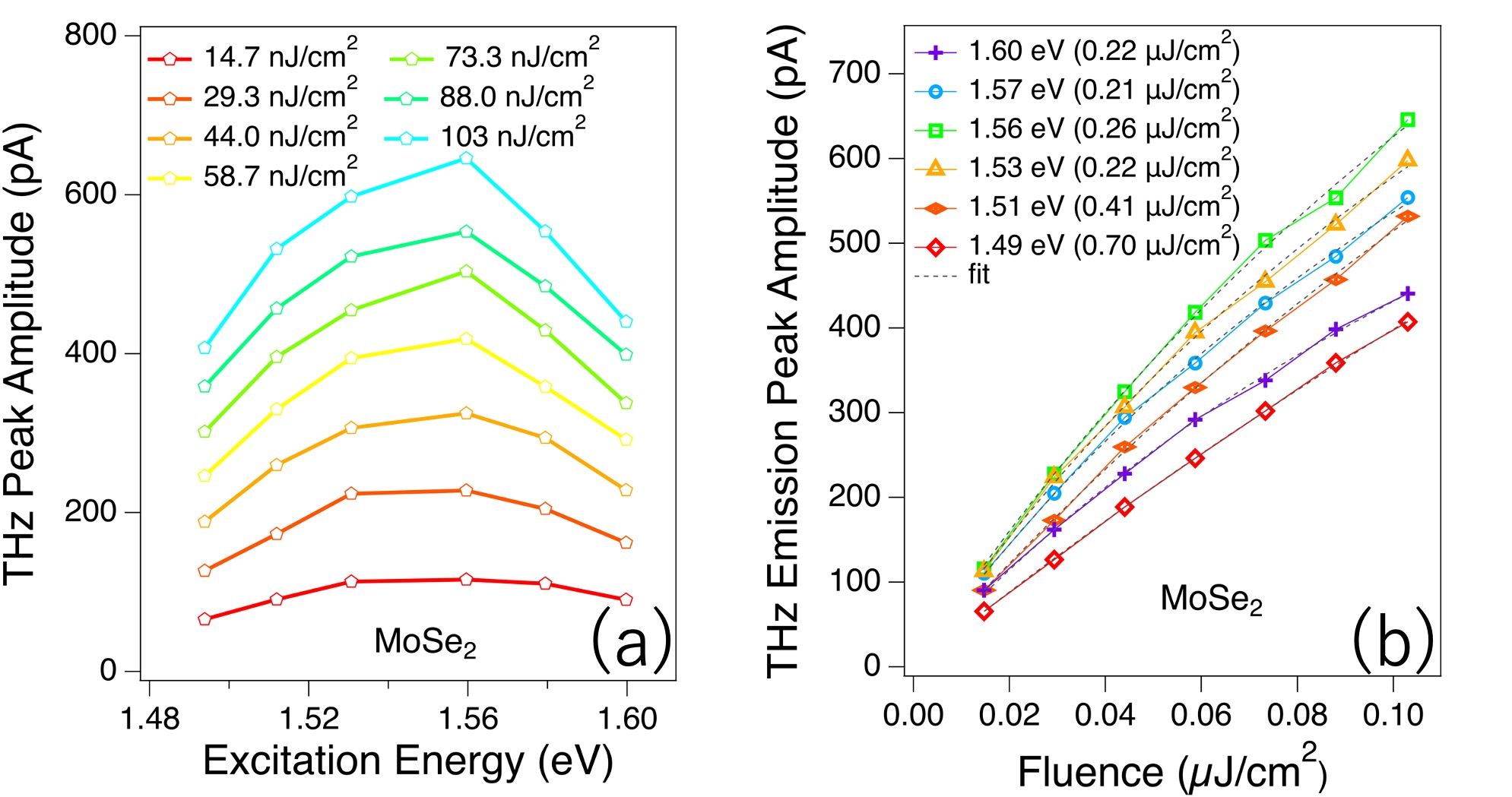}%
\caption{\label{fig:kpoint} (a) The transient-current driven THz peak amplitude (peak of $E_{\text{THz,curr}}$) plotted against excitation energy. Maximum can be found at $\sim$1.56 eV, in the vicinity of the $A$-gap in the MoSe$_2$ bandstructure's K-point. The behavior is more prominent as the incident fluence increases. (b) The THz peak amplitude also increases with fluence, but shows signs of possible saturation. Fitting with a saturation equation, the saturation fluence is indicated for every excitation energy. }
\end{figure}

We also confirmed the relationship between the THz amplitude and the pump wavelength and fluence. The wavelength-dependence of the peak THz amplitude (time delay = 10 ps) for MoSe$_2$ is shown in Fig.3(a), where the wavelength is varied between 775 nm (1.60 eV) to 830 nm (1.49 eV). The peak THz amplitude is highest when pumped at 795 nm (1.56 eV), which is near the estimated $A$-gap found at the K-point of the MoSe$_2$ band structure \cite{nevinskas2021terahertz,haldar2015systematic,yang2022first}. This effect is emphasized at higher incident fluence. When pumping at electron resonances, higher carrier mobility is expected, as well as efficient carrier excitation. The high mobility and increased photocarrier population amplifies the THz radiation.  Additionally, the fluence-dependence of the peak amplitude is plotted in Fig.\ref{fig:kpoint}(b), with the THz emission increasing with fluence. The curves were fitted using a saturation equation $C/\left( 1+F_{\text{sat}} / F \right)$, where C is an amplitude fitting factor, F is the fluence and $F_{\text{sat}}$ is the saturation fluence. For THz emission driven by transient currents, saturation occurs as an effect of Coulomb screening by photogenerated carriers \cite{Benicewicz:94}. The values of the saturation fluence are shown in the figure. Results reveal that the incident laser fluences used in this experiment are an order of magnitude below saturation. 

The results of the azimuthal angle dependence, the fluence- and wavelength- dependence measurements confirm that the initial cycle emission predominantly originates from photo-induced transient currents, \begin{equation}   E_{{\text{THz},curr}}\propto\frac{\partial J}{\partial t}
\end{equation}
We provide a simplified current equation that sufficiently fits our data,
\begin{eqnarray}
    J(t)&&= en(t)v(t)\nonumber \\ 
&&\propto\text{erf} \left(\frac{t-t_0}{\tau} + 1\right) \text{exp}\left(-\frac{t-t_0}{\alpha}\right) \text{exp}\left(-\frac{t-t_0}{\beta}\right)
\end{eqnarray}

The transient current in Eq.(3) is proportional to the carrier population $n(t)$ created by the arrival of a Gaussian laser pulse on the surface at time $t_0$, and moving with velocity $v(t)$. We simplified the expression to contain the effective temporal resolution of the system in $\tau$, and the carrier population falls off with time constants $\alpha$ and $\beta$, which reflect the effect of the material’s carrier lifetime. 

The oscillations are described by a sum of sinusoidal functions{\cite{hase1996optical},
\begin{equation}
E_{{\text{THz},ph}}\propto \sum_{i=1,2}B_{i}\text{exp}\left(-\frac{t-t_{ph}}{\tau_{ph,i}}\right)\sin(2\pi f_i t +\phi_i)  
\end{equation}
where $B_i$ represent the amplitudes, $\tau_{ph,i}$ correspond to the damping constants, $f_i$ are the oscillation frequencies, and $\phi_i$ describes their phases. The oscillations appear at a time delay $t=t_{ph}$, shortly after the onset of $E_{\text{THz,curr}}$ at $t=t_0$. The total waveforms are the superposition of the current-driven THz emission and the coherent phonon radiation, 
\begin{equation}
E_{\text{THz}}=E_{{\text{THz},curr}}+E_{\text{THz},ph} 
\end{equation}
\begin{widetext}
\center{\begin{table}
\caption{\label{Tab:parameters} Fitting parameters of the interlayer phonon emission. MoSe$_2$ data was obtained under 795 nm excitation, and WSe$_2$ data was obtained under 765 nm excitation.}
\begin{tabular}{|l|l|l|l|l|l|l|l|l|l|}
\hline
 & t$_{ph}$ (ps) & $\tau_{ph,1}$ (ps) & f$_1$ (THz) & $\phi_1$ (rad) & $\tau_{ph,2} $ (ps) & $f_2$ (THz) & $\phi_2$ (rad) \\ \hline
MoSe$_2$ & 11.18 $\pm$ 0.032 & 0.490 $\pm$ 0.016  & 0.357 $\pm$ 0.150 & 0.391 $\pm$ 0.099 & 3.965 $\pm$ 0.845 & 0.599 $\pm$ 0.151   & 0.466 $\pm$ 0.408 \\ \hline
\begin{tabular}[c]{@{}l@{}}WSe$_2$ \end{tabular} & 11.33 $\pm$ 0.028 & 1.05 $\pm$ 0.142 & 0.176 $\pm$ 0.058 & 1.013 $\pm$ 0.013 & 0.447 $\pm$ 0.053 & 0.982 $\pm$ 0.018 & 0.662 $\pm$ 0.063 \\ \hline
\end{tabular}
\end{table}}
\end{widetext}
The time-domain data runs from $t=0$ to 35 ps, with the initial pulse peak centered at $t=10$ ps. Waveform fitting was conducted for the THz emission signal found between $t = 5$ to $25$ ps, and is overlaid with the data in Fig.1(a). To create the waveform fit $E_{THz,fit}$, the software used replaces the points outside of $5\text{ ps}<t<25\text{ ps}$ with the experimental data. The parameters obtained from the full waveform fit, Eq. (5), allows us to separate the two contributions $E_{THz,curr}$ and $E_{THz,ph}$, shown with y-offsets. Parameters used for the fitting curves are detailed in Table I. The phonon response appears at the tail end of the current-driven THz emission, around $t_{ph} \sim 11.18$ ps. The best waveform fits that accurately reproduce not only the time-domain waveform, but also the power spectra, could be obtained by using two sinusoidal functions for $E_{THz,ph}$. Table \ref{Tab:parameters} gives the details of the fitting parameters for the phonon modes, as in Eq.(4). The frequencies obtained are $f_1=0.357$ THz (11.90 cm$^{-1}$) and $f_2=0.599$ THz (19.98 cm$^{-1}$), which have damping constants of $\tau_{ph,1}=0.490$ ps and $\tau_{ph,2}=3.965$ ps, respectively. Coherent lattice vibrations that emit terahertz radiation are IR-active\cite{dekorsy1995emission,tani1998terahertz}. The IR-activity of the interlayer vibrations depend on the number of layers participating, and the resulting symmetry of the vibrations\cite{zhang2015phonon}. Exact assignment of the observed modes require rigorous analysis of the layer symmetry and is beyond the scope of this work. Nonetheless, based on available reports\cite{zhang2015phonon,liang2017low,zhao2013interlayer}, the lower frequency mode $f_1$ is possibly a layer-breathing mode. And $f_2$ is likely an interlayer shear mode, similar to what has been reported by Liang, et al\cite{liang2017low} for few layer MoSe$_2$ at 19 cm$^{-1}$. Differences in reported values may be attributed to the behavior of the vibrational modes as the number of participating layers increase - shear modes redshift, while layer breathing modes blueshift\cite{liang2017low}.

The Fourier transform for $E_{THz,curr}$ (blue dashed curve in Fig.2(b)) is included in the plot to show the broad distribution of frequencies, centered at $\sim$0.4 THz  and extending to 2.5 THz, which is typical for single-cycle pulse THz emission. It also serves as a visual contrast against the MoSe$_2$ full waveform's frequency spectra. The difference in spectral features of the total emission between the two conditions stems from the temporal overlap between the main THz pulse and the phonon emission. The presence of phonon modes create prominent peaks (as well as dips) in the frequency spectra. The modes obtained from the fitting do not necessarily coincide with these peaks or dips, but rather, the peaks and dips indicate positive (as well as negative) interference between $E_{\text{THz,curr}}$ and $E_{\text{THz,ph}}$. 

THz emission from pristine WSe$_2$, which shares the same crystal structure as MoSe$_2$, also showed similar features. The time-domain waveform and frequency spectra are shown in Fig.\ref{fig:wse2}, together with the results of the fitting. Fig.4(c) shows the wavelength-dependence of the $E_{THz,curr}$ emission peak amplitude. The maximum amplitude value was obtained under an excitation wavelength of 765 nm (1.62 eV). This is consistent with the approximate energy of the $A$-gap found in the WSe$_2$ band structure's K-point. The THz emission waveform also shows a single cycle pulse followed by damped oscillations that begin at around $t_{ph}\sim11.33$ ps. We obtained the best fit using two sinusoidal functions with frequencies of $f_1=0.176$ THz (5.870 cm$^{-1}$) and $f_2=0.982$ THz (32.75 cm$^{-1}$). Fitting parameters are listed in Table \ref{Tab:parameters}. The phonon oscillations decay within $\tau_{ph,1}\sim$  1.050 ps and $\tau_{ph,2}\sim$  0.447 ps. These decay constants are consistent with a recent report of 2D WSe$_2$ phonon lifetimes, estimated to be close to 1 ps, obtained by analyzing the power-dependence of Raman scattering data\cite{bandyopadhyay2022spectroscopic}.  The lower frequency mode $f_1$ is likely to be a layer breathing mode\cite{zhang2015phonon}. For the possible vibration mode associated with $f_2$, a bulk shear mode at around 23.8 cm$^{-1}$ has been reported\cite{liang2017low,zhao2013interlayer} for WSe$_2$ using linear chain model (LCM) calculations, and a minor breathing mode at 32.4 cm$^{-1}$ was also reported from DFT calculations\cite{zhao2013interlayer} of an 8-layer crystal. 

The THz emission from WSe$_2$ also remained p-polarized under pump-polarization-dependence measurements (not shown), indicating that transient current effect is also dominant for this sample. Comparing the phonon oscillation amplitudes between samples, the amplitude of the coherent phonon emission, $E_{\text{THz,ph}}$, relative to $E_{\text{THz,curr}}$ is higher in WSe$_2$. 

\begin{figure}
\includegraphics[width=0.5\columnwidth]{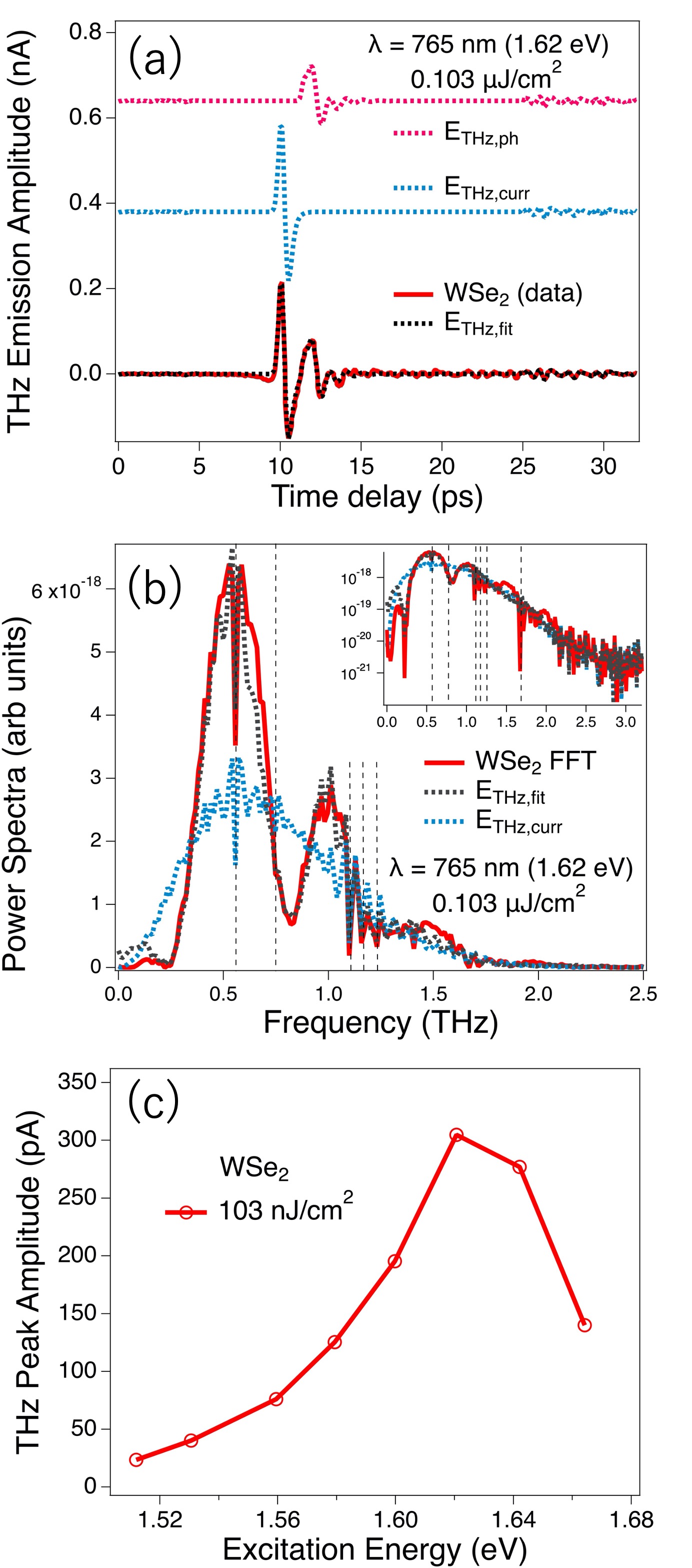}%
\caption{\label{fig:wse2}THz emission from WSe$_2$ (a) time-domain waveform, along with the results of waveform fitting (b) frequency spectra of the emission data, the full waveform fit, and the single cycle pulse; and (c) wavelength dependence of $E_{THz,curr}$ pulse's peak amplitude. }
\end{figure}

When THz emission from coherent phonons were first detected, they were from materials such as Te, GaAs, or Ge which have IR-active LO and TO phonons\cite{tani1998terahertz,dekorsy1995emission}. In a recent report\cite{guzelturk2018terahertz}, THz emission from hybrid perovskite methylammonium lead iodide (CH$_3$NH$_3$PbI$_3$) showed contributions from several coherent phonon modes – the LO phonon mode at 1.2 THz (40 cm$^{-1}$), a coupled vibration mode at 2.8 THz (93.4 cm$^{-1}$) due to the presence of organic and inorganic sublattices, as well as the photo-Dember field-induced coherent vibration of the lattice at 0.65 THz (21.7 cm$^{-1}$). The observed THz emission from the hybrid perovskites is quite visually similar to our MoSe$_2$ and WSe$_2$ data wherein a single cycle pulse appears, followed by a short oscillatory burst. In a report on the low-frequency Raman scattering of WSe$_2$\cite{zhao2013interlayer}, the Raman-active layer-breathing mode is shown to shift from $\sim$28 cm$^{-1}$ for a single layer of WSe$_2$ to 8 cm$^{-1}$ for a 7-layer sample. It can be inferred that as the number of layers increase, mode frequency may be beyond the limit of their apparatus (scale is only up to 5 cm$^{-1}$). This, and other reports show the possibility of interlayer vibrations having such low frequencies; and underline that these modes may either be Raman-inactive or that these modes cannot be detected by ordinary back-scattering set-ups\cite{liang2017low,ge2014coherent}. For THz emission spectroscopy, the detection sensitivity is highest around 17-33 cm$^{-1}$ (0.5-1 THz)\cite{Kamo2014,jepsen1996generation}, with a lower frequency limit of around 5 cm$^{-1}$ (0.15 THz) for typical LT-GaAs dipole photoconductive antennas. While the exact assignment of the vibrational modes found for our TMDCs require more rigorous computational methods, the possibility to observe IR-active low frequency vibrational modes in layered materials through THz emission spectroscopy has been demonstrated. Additionally, THz emission spectroscopy offers ease-of-use compared to THz-TDS in that it does not require cumbersome sample preparation, and has a finer spatial resolution in the micrometer range. 

Analysis of the THz generation mechanism in single crystal 2H-MoSe$_2$ and 2H-WSe$_2$ shows contributions from transient currents and coherent interlayer phonon vibrations. Through the waveform analysis of the time-domain THz emission, the frequencies of the interlayer vibration modes have been estimated, revealing coherent phonon modes as low as $\sim$5.87 cm$^{-1}$. The clear observation of these vibrational modes demonstrated that THz time-domain emission spectroscopy is a powerful tool to observe and analyze low-frequency phonon modes in TMDCs that are non- or weak-Raman active. 

%
%

%

\begin{acknowledgments}
This work was supported in parts by grants from CREST-JST (Grant Number JPMJCR1875), Japan and the Cooperative Research Program of Research Center for Development of Far Infrared Region, University of Fukui (R04FIRDG032A). JA also thanks Dr. Naoki Yamamoto (Jichi Medical University) for discussions on THz-based vibrational spectroscopy, and Dr. Soliman Garcia (Kyoto University) for discussions on wave fitting, and technical assistance.
\end{acknowledgments}

\bibliography{references}

\end{document}